\def\deg{^{\circ}}
\def\etal{{\it et al.}\thinspace}

\def\eg{{\it e.g.,}\thinspace}
\documentclass[useAMS,usenatbib]{mn2e}
\usepackage{graphicx}

\title[Core and Conal Component Analysis of Pulsar B1237+25]{Core and Conal Component Analysis of Pulsar B1237+25}
\author[Zuzana Srostlik and Joanna M. Rankin ]
{Zuzana Srostlik$^{1}$\thanks{zuzana.srostlik@uvm.edu; Joanna.Rankin@uvm.edu} and Joanna M. Rankin$^{2,1}$\footnotemark[1]\\
$^{1}$Physics Department, 405 Cook Physical Science Building, University of Vermont, Burlington, 05405, USA\\
$^{2}$Sterrenkundig Instituut `Anton Pannekoek', Universiteit van Amsterdam, Kruislaan 403, SJ 1098 Amsterdam, Netherlands}

\begin{document}

\date{Accepted 2004 month day. Received 2004 month dat; in original form 2004 month day}

\pagerange{\pageref{firstpage}--\pageref{lastpage}} \pubyear{2004}

\maketitle

\label{firstpage}

\begin{abstract}
The paper provides a new analysis of this famous five-component ({\bf M}) pulsar.  In addition to the star's core-active ``abnormal'' mode, we find two distinct behaviors within its ``normal'' mode, a  ``quiet-normal'' mode with regular 2.8-period subpulse modulation and little or no core activity, and a ``flare-normal'' mode, where the core is regularly bright and a nearly 4-period modulation is maintained.  The ``flare-normal'' mode appears to be an intermediate state between the ``quiet normal'' and ``abnormal'' behaviors.  Short 5--15-pulse ``flare-normal''-mode ``bursts'' and ``quiet normal'' intervals alternate with each other quasi-periodically, making a cycle some 60--80 pulses in duration.   ``Abnormal''-mode intervals are interspersed within this overall cycle, usually persisting for only a few pulses, but occasionally lasting for scores or even many hundreds of pulses.  

Within subsequences where the core is exceptionally quiet, the pulsar provides a nearly ``textbook'' example of a central PA sightline traverse---with a PA rate measured to be at least some 180$\deg/\deg$.  On this basis it is shown that the sightline impact angle $\beta$ must be about 0.25\degr\ or some 5\% of the outer conal beam radius.   

The star's core component is found to be incomplete, despite the fact that the core's full antisymmetric circularly polarized signature is present.  The visible core component aligns with the trailing right hand (RH) portion of the circular signature.  Measures either of the circular signature or of the the trailing half of the core, however, reiterate the roughly 2.6\degr\ angular diameter of the star's polar cap.  

We find that the star's PA traverse is disrupted through the action of orthogonally polarized linear power in the longitude range of the core component.  This circumstance can be seen in the  ``hook'' under the core component which usually entails {\em four} sense reversals of the PA, with its center falling at the same PA as the extrema of the overall traverse.  This behavior is modeled to show its effect.  

Finally, the star provides two well defined fiducial points from which emission-height estimates can be computed, the respective centers of both the linear PA traverse and the zero-crossing point of the circular polarization signature.  We thus find that the outer and inner cones are emitted at heights of some 340$\pm79$ and 278$\pm$76 km, respectively, such that their fieldline ``feet'' are some 78$\pm$7 and 53$\pm$5\% of the polar cap radius.  We also find evidence that the core emission occurs at a height of some 60 km.  
\end{abstract}


\section{Introduction}
Pulsar B1237+25 has been widely studied and represents the classic example of a pulsar with five emission components, apparently produced by a sightline traverse which passes almost through the center of two concentric emission cones and a central core beam [\eg \citet{b11}]. It is also known to exhibit a ``normal'' and ``abnormal'' emission mode, frequent ``null'' pulses, and regular 2.8-period/cycle subpulse modulation [Backer (1970a-c,1973)] in its ``normal'' mode.  Each of these properties has been well investigated, but little is known about possible linkages between them.  We have carried out a new study of B1237+25's polarized pulse sequences (hereafter PSs) with the purpose of investigating a) the characteristics and activity of its intermittently active core component and b) the recent claim that the star exhibits a third weak inner cone [\citet{b3}].  The pulsar's individual-pulse behavior is so complex and varied, particularly with the benefit of excellent new Arecibo 327-MHz observations, that we have had to question again what interpretation should be made of its five profile features.\footnote{Indeed, some early profiles suggest six components, ostensibly due to use of one circular or linear feed.  This   understanding has persisted, though, even with better observations (\eg \citet{b9}; \citet{b21}) and we will develop some basis for understanding why below.} PS polarization has provided the foundation for our analyses.  Following the published studies of total sequences at 430 MHz [\citet{b12}] and 1400 MHz [\citet{b5}], \citet{b9} investigated the specific characteristics of the star's two modes.  Of particular significance in our current context was the discovery that the ``normal'' and ``abnormal'' emission (or profile) modes represented differing proportions of primary- and secondary-polarization-mode (hereafter PPM and SPM) power.  We also draw on the studies of \citet{b6}, \citet{b7}, \citet{b33} and \citet{b10}.  

While B1237+25's relatively weak core component (III) has influenced beaming models [\eg \citet{b13}], its actual characteristics remain somewhat obscure.  Its steeper spectrum renders it ever weaker at higher frequencies, and even around 400 MHz its close proximity to the stronger trailing component (IV) makes it difficult to isolate for study.  We therefore began our study by focussing on the population of single pulses where the core was active.  We quickly relearned that the core is active episodically, 
continuously so in ``abnormal''-mode sequences, but also strongly at fairly regular intervals within ``normal''-mode sequences.  We thus now conclude that the classical ``normal'' mode is actually an alternation of two modes, a ``quiet normal'' and a ``flare-normal'' mode.  

\S 2 then describes our observations, \S 3 \& 4 introduce the characteristics of the three modes.  In \S 5 we present the results of our search for a third cone following \citet{b3}. \S 6 gives an analysis of the star's nulling behavior, and \S 7 explores how the three modes interact.  In \S 8 \& \S 9 we discuss our results regarding the geometry and polarization of B1237+25's core component.  \S 10 gives an analysis of the star's emission height, and \S 11 provides a summary and gives our overall conclusions.

\section[]{Observations}
The observations used in our analyses were made using the 305-m Arecibo Telescope in Puerto Rico.  The primary 327-MHz polarized PSs were acquired using the upgraded instrument together with the Wideband Arecibo Pulsar Processor (WAPP\footnote{http://www.naic.edu/$\sim$wapp}) on 2003 July 13, 14 and 21, comprising 2340, 5094 and 4542 pulses, respectively.  The ACFs and CCFs of the channel voltages produced by receivers connected to orthogonal linearly polarized feeds were 3-level sampled.  Upon Fourier transforming, 64 channels were synthesized across a 25-MHz bandpass with a 512-$\mu$s sampling time, providing a resolution of 0.133$\deg$ longitude.  The Stokes parameters have been corrected for dispersion, interstellar Faraday rotation, and various instrumental polarization effects.  Similar observations in four 21-cm bands on 2003 August 4 used 100-MHz widths, recorded 2062 pulses, and were sampled at the same resolution.  Older Arecibo observations at 430 and 111.5 MHz are used for comparison as noted below.  

\begin{figure} 
\includegraphics[width=80mm]{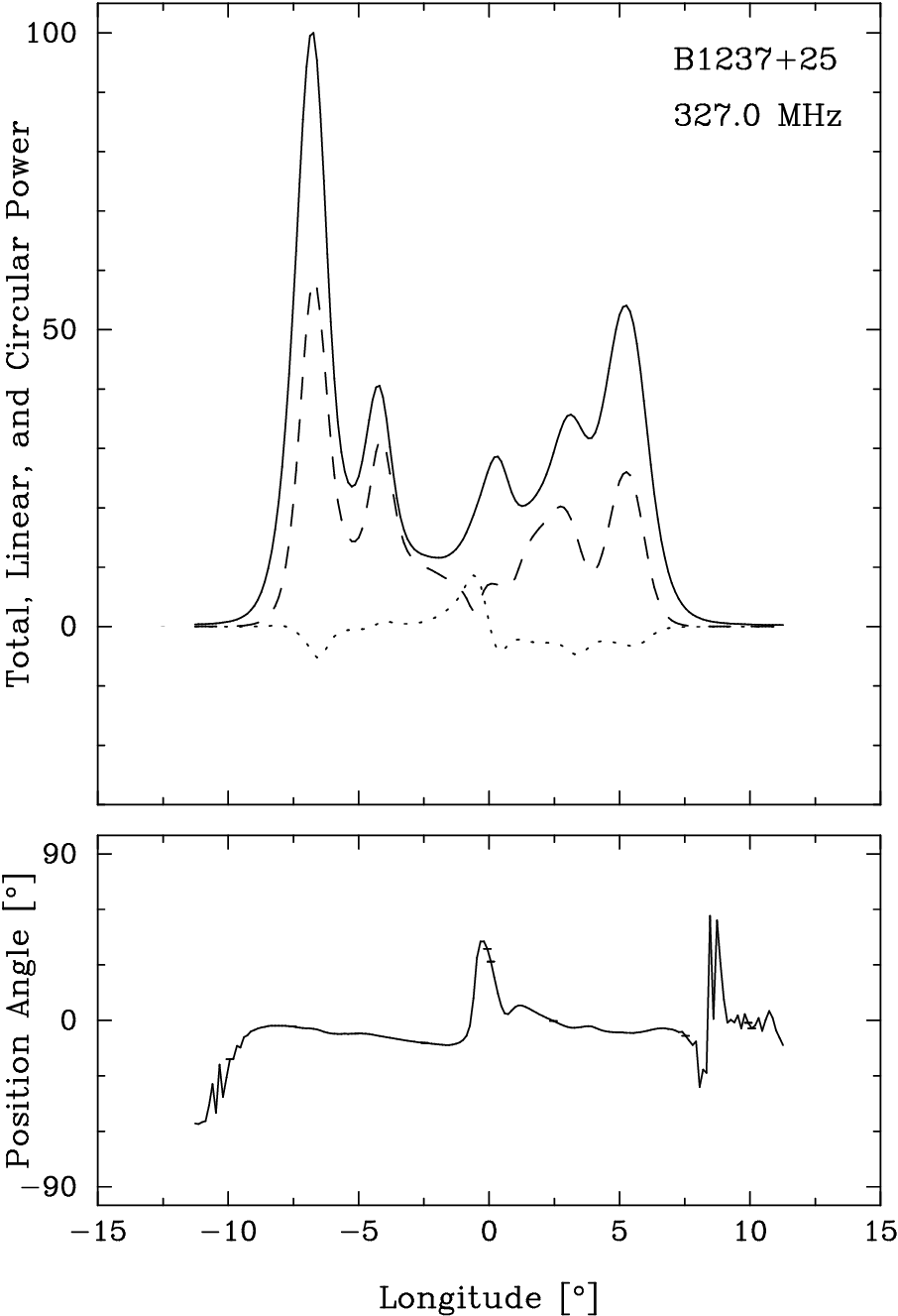}
\caption{Total-average profile of the 327-MHz PS recorded on 2003 July 21.  Here, no effort is made to discriminate between the star's modes or behaviors.  The total-intensity Stokes parameter $I$, linear polarization $L$ [=$(Q^2+U^2)^{1\over 2}$] and circular polarization $V$ (=LH--RH) are plotted in the upper panel (solid, dashed and dotted curves, respectively), and the polarization angle (hereafter PA) $\chi$ [=${1\over 2}\tan^{-1} U/Q$] is given in the lower one.  Here, and in the following figures, the longitude origin has been taken close to PA-traverse inflection point (see text).  We see the classic five-component average profile with its nearly constant PA apart from ``jumps'' on the profile outer edges and near the center.  The outside features result from orthogonal-polarization-mode (OPM) dominance changes, whereas the origin of the central ``hook''---where we might expect a steep, nearly 180$\deg$ PA rotation---remains something of a mystery which we will attempt to explicate below. Note the pronounced linear depolarization at this longitude and the weak left- to right-hand $V$ signature near the core (comp. III) longitude.}
\label{Fig1}
\end{figure}

\begin{figure} 
\includegraphics[width=80mm]{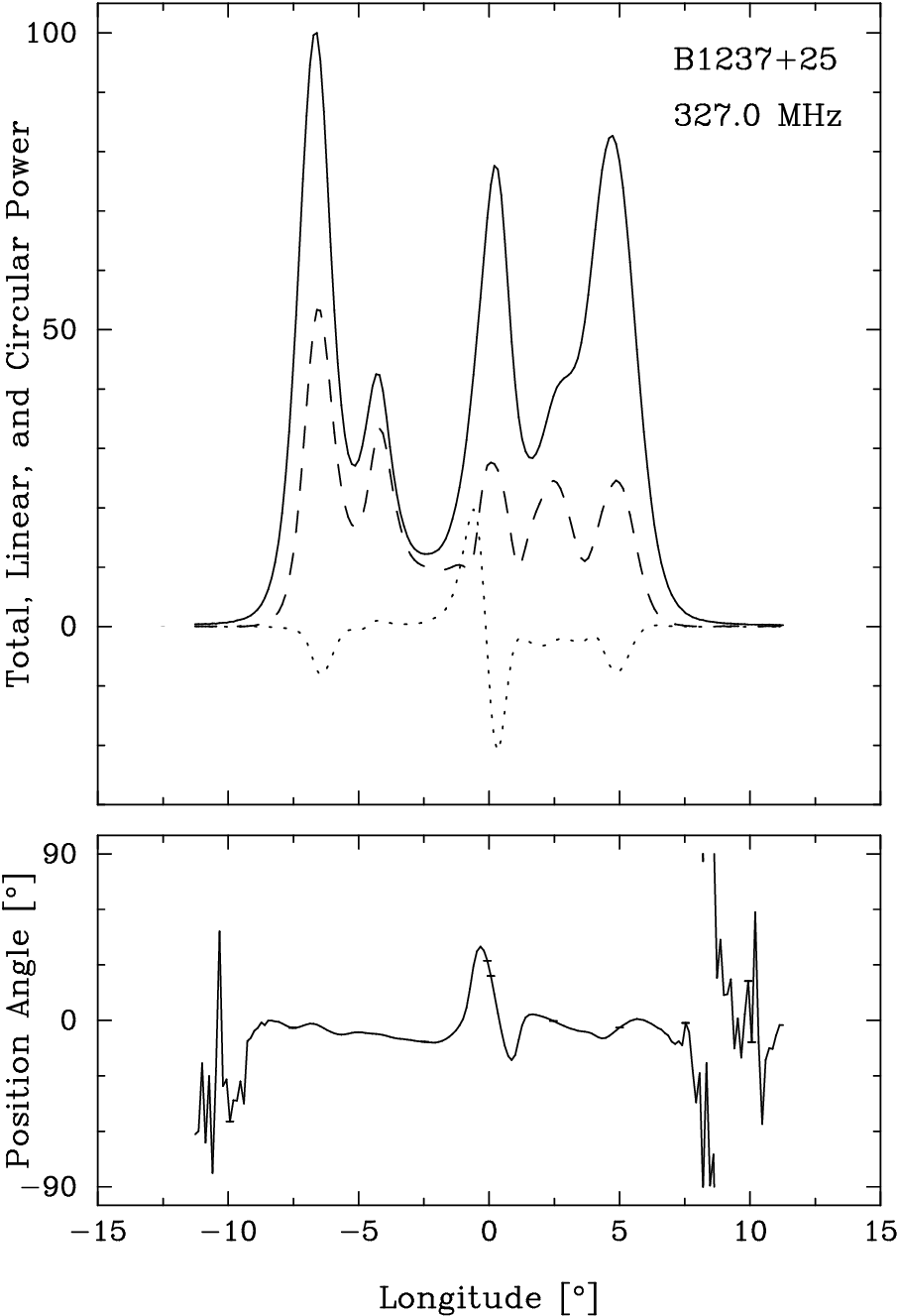}
\caption{Average profile of ``flare-normal''-mode intervals at 327 MHz on 2003 July 21.  Note the brighter,  more highly polarized core component with its antisymmetric circular signature.  The trailing component becomes brighter and less linearly polarized and is seen to partially overlap comp. IV.   We note also that the deep minimum in emission between components II and III persists in the ``flare-normal'' mode.  This profile is comprised of 370 pulses, representing some 50 ``flare-normal''-mode PSs of typically 5--10 pulses duration.}
\label{Fig2}
\end{figure}

\begin{figure} 
\includegraphics[width=80mm]{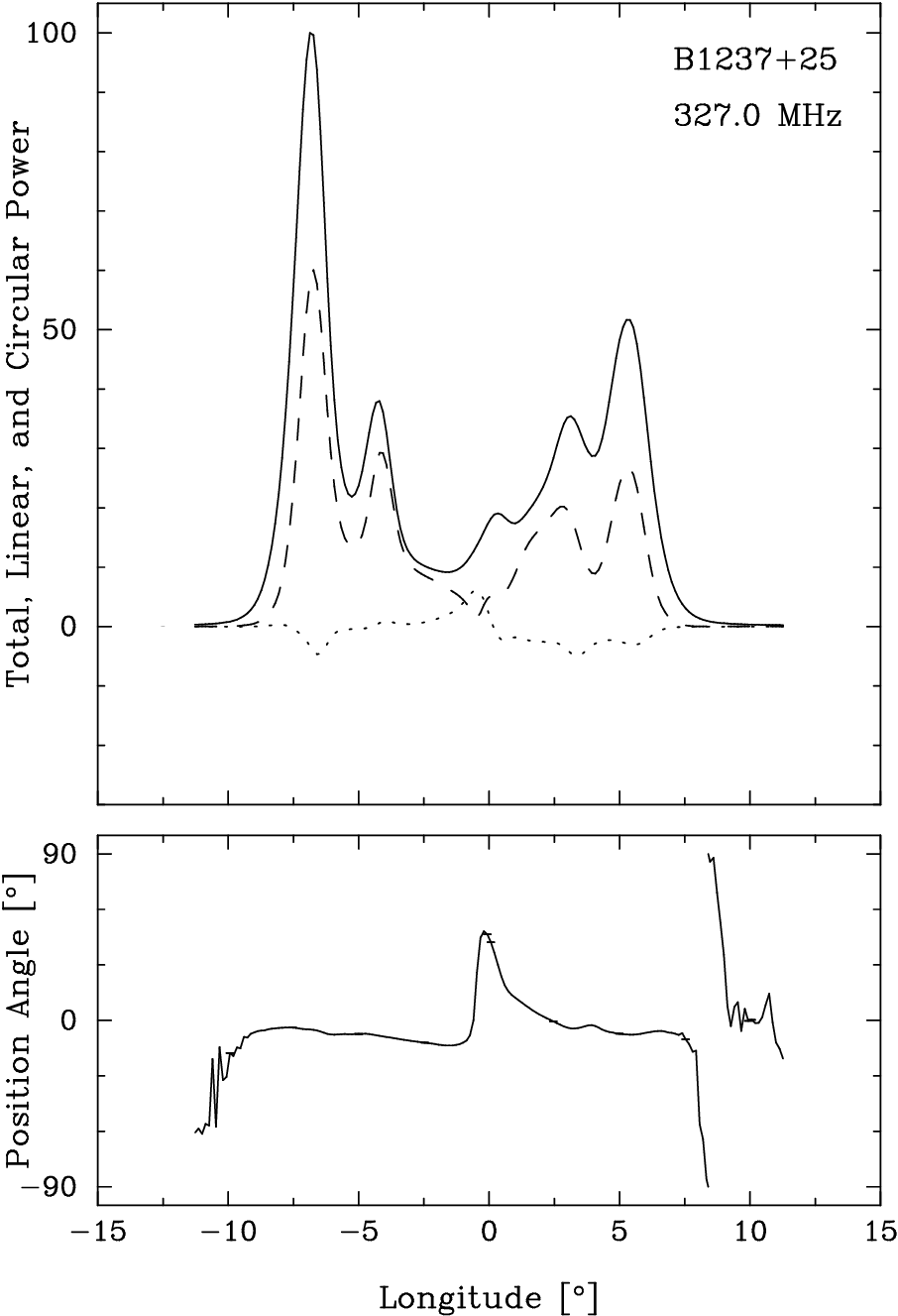}
\caption{Average profile of ``quiet normal''-mode sequences at 327 MHz on 2003 July 21.  Core activity in these intervals is weak, indeed virtually undetectable in many of the constituent single pulses.   This profile represents the star's behavior between ``flare-normal''- and rarer ``abnormal''-mode intervals.  Note how there are five peaks, II and IV being about the same height, I being significantly more intense than V, and III very weak.  Notice also the slightly different shape of the central PA feature; many of the individual pulses exhibit a barely resolved, negative, 180\degr\ excursion which is here less affected by core emission at the same longitude.  The diagram represents the average of 2654 pulses which were identified as belonging to the ``quiet normal'' mode.}
\label{Fig3}
\end{figure}

\begin{figure} 
\includegraphics[width=80mm]{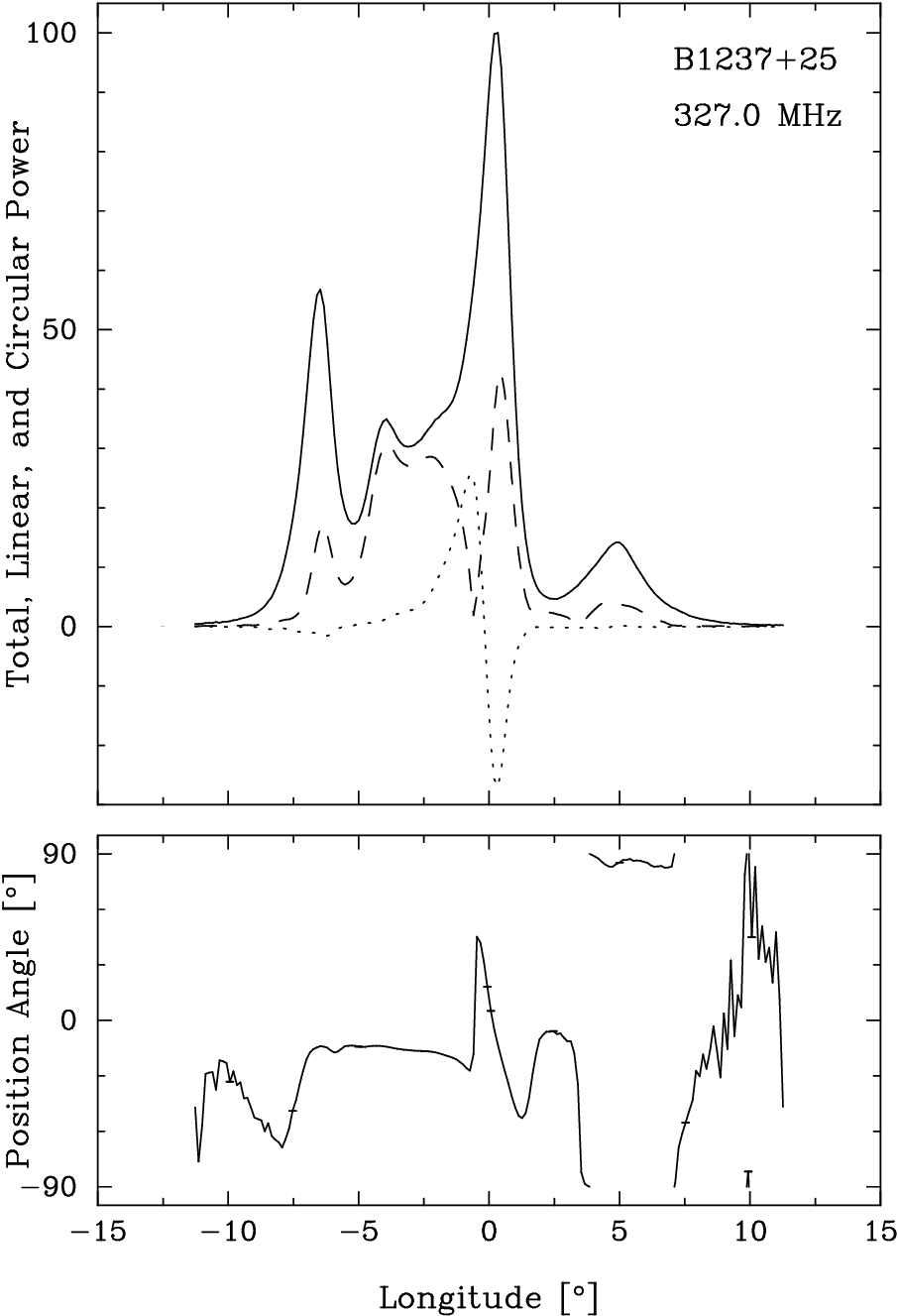}
\caption{Partial-average profile of the ``abnormal''-mode intervals at 327 MHz on 2003 July 21.  Notice the bright core component with its strongly linearly polarized trailing edge and antisymmetric circular signature.  Note also the linear polarization minimum at very nearly the same longitude as in the total and ``quiet normal'' profiles---but here the fractional linear polarization immediately before and after it is quite high.  Here, the core properties contrast very strongly with those of the ``flare-normal'' mode.  The ``abnormal''-mode profile exhibits a dramatic rise in the intensity of the ``bridge'' region between components II and III, where the fractional linear polarization is always large and thus nearly (PPM) unimodal.  Finally, the PA traverse under the trailing portion of the core component is especially well defined here and represents an about 90\degr\, negative-going traverse of SPM power (see text).  Note also that the overall width of the ``abnormal''-mode profile has decreased slightly and its depolarized edges and accompanying PA features show the effects of increased levels of SPM power.  The profile is comprised of 91 pulses.}
\label{Fig4}
\end{figure}

\section[]{Integrated Profile Analysis}
Figure~\ref{Fig1} reiterates for ease of comparison the familiar total profile of B1237+25, created by averaging a set of pulses without distinguishing its modes.  We see the five classical components (I-V), which have usually been interpreted as a central sightline traverse through two concentric emission cones and a core beam.  The validity of this inference, based on simple average profiles like this one, must be questioned for many reasons.  The profile form is far from symmetric in either total power (solid curve) or linear polarization (dashed curve), and we see neither the full antisymmetric circular signature (dotted curve) often associated with a core component nor the expected 180\degr-position angle (PA) excursion of a central sightline traverse. Rather, the core region is depolarized as clearly shown by the linear minimum.  Indeed, one might conclude almost immediately from such a profile alone that it cannot be comprised of a single orthogonal-polarization-mode (hereafter OPM) population of individual pulses.  We know from the papers quoted above that both a ``normal'' and ``abnormal'' mode have been identified and well studied in this pulsar---and moreover that they differ considerably in relative OPM power.  We argue below that the classical ``normal'' mode in fact consists of an alternation of two new modes, a ``flare-normal'' mode characterized by intervals of strong core activity, and a ``quiet normal'' mode wherein the core emission is nearly undetectable.  

In our attempt to identify populations of individual pulses in which the core component is particularly active, we found that such pulses occur either in ``abnormal''-mode intervals or within fairly discrete ``normal''-mode intervals.  Further investigation showed us that these bright-core intervals typically persist for 5-10 pulses and that they recur quasi-periodically about every 60-80 periods (hereafter $P_1$).  We thus begin by delineating the characteristics of this ``flare-normal'', bright-core mode in relation to the ``quiet normal'' and classical ``abnormal'' modes.  

\subsection[]{Third Mode}
Figure~\ref{Fig2} represents the average profile of the ``flare-normal''-mode PSs, identified by visual inspection of the full observations using colour displays such as that in Figure~\ref{Fig5}.  As we will show below, the boundaries of ``flare-normal'' mode subsequences can be distinguished from both ``abnormal''- and ``quiet normal''-mode intervals in these high quality observations generally within a pulse or two.  As expected, the ``flare-normal''-mode profile shows a much brighter core component, and we see the antisymmetic circular polarization often associated with core components.  Note also that the core component is no longer linearly depolarized as in the total profile or classical ``normal'' mode.  It here exhibits a symmetrical $>$25\% linear ``component'' just under the total-power core and this is accompanied by a strong negative excursion of the PA (lower panel).  

Components I, II and IV appear almost unaltered in ``flare-normal''-mode sequences, but comp. V increases markedly in intensity and appears to occur slightly earlier, thus partially overlapping comp. IV.  This modal component V is also substantially less linearly polarized, showing the effect (as we will see below) of an increased level of SPM power.  The subtle decrease in the spacing of components I and II also appears significant and reflects a slightly narrower overall profile width as compared with the total profile or the ``quiet normal''-mode profile below.

\subsection[]{``Quiet normal'' Mode}
Figure~\ref{Fig3} gives an average profile for the ``quiet normal''-mode sequences.  It represents the sum of those intervals exhibiting little or no core activity and in practice it differs indistinguishably from an average of the residuum after ``flare-normal''  and ``abnormal''-mode intervals have been identified and removed.  Apart from the decreased core presence, this profile differs little from the typical total profile, because some 85\% of ``normal''-mode pulses are ``quiet normal'' and the ``abnormal'' mode is yet rarer.\footnote{Occasional ``abnormal''-mode-dominated total profiles do occur, however, because it can sometimes persist for hundreds, or perhaps even thousands, of pulses.} We again see a sharp minimum of the linear polarization near the longitude origin, and perhaps (owing to less core influence) a somewhat larger portion of the mostly unresolved negative PA excursion near the profile center.  It is also noteworthy that in individual pulses comp. II is usually very narrow (a few bins) and is 100\% polarized in typically 90\% of its occurrences.  The ``quiet normal''-mode PA very often exhibits a full negative, nearly 180\degr, excursion under the core component.  We see this behavior in many individual pulses, but not always in modal average profiles.  Weak core power---often hardly detectable in individual ``quiet normal'' pulses---at times still accrues to form a distinct profile component; and when present, this residual core emission is often sufficient to disrupt the ``conal'' PA traverse.  The effect is most pronounced at the linear power minimum in the profile center which indeed coincides with the positive-going PA ``jump'' under it.  Thus, in many individual pulses and modal averages, when the core power is low enough, this ``jump'' connects negatively and symmetrically.

\subsection[]{Classical ``abnormal'' Mode}
Figure~\ref{Fig4} gives the average profile of a long, ``abnormal''-mode sequence and, in using the word ``classical'', we emphasize that our criteria for identifying pulses belonging to this mode are the same as those in published studies.  Again, the profile alterations exhibited by the pulsar's ``abnormal'' mode are so dramatic at meter wavelengths that at first sight we can wonder if we are observing the same pulsar.  Each of the five components assumes a different relative configuration in the ``abnormal'' mode, and the emission minimum, usually seen just before the core component, is now observed to follow it.  The linear-polarization minimum, however, is even more prominent here, and falls at almost exactly the same longitude as in the ``quiet normal'' and total profiles.  Also, the here positive ``jump'' associated with the minimum connects negatively in other such modal profiles, resulting in an overall PA traverse of more than -210\degr!  Something of this behavior can also be seen in Bartel \etal's (1982) fig. 7 (left).  

The core component, of course, predominates in the ``abnormal'' mode and we will see below that it is almost {\em continuously} active in adjacent individual pulses.  Note, however, that while the circularly polarized signature of the core component is perhaps strongest here, the linear polarization properties of the ``abnormal''-mode core are very different than those of the ``flare-normal'' mode.  Here, we see even larger fractional linear polarization and again a long (but here longer) negative PA traverse associated with it.  Both the total intensity and linear peaks fall slightly later than in the ``flare-normal'' mode, such that the linear polarization appears to trail within the visible core component.  Moreover, this linear is delineated by leading and trailing minima which coincide closely with the duration of the negative (RH) part of the circular signature.  All this then leaves the leading part of the visible core component conspicuously linearly depolarized.  

Note also that the overall modal profile width is again narrower, by nearly 2\degr, than the total profile.  Components IV and V have merged and highly polarized emission is now found in the ``empty'' region between components II and the core.  Over much of the width of the profile we see evidence of increased SPM power: the PA of comp. V is now SPM dominated as is a leading-edge region of comp. I, but also we see evidence of PPM power on the extreme outer edges.  Only the regions under and following comp. II and just after the core remain PPM dominated---this in stark contrast to the ``normal'' or total profiles where the PPM dominates ``conal'' regions of the profile everywhere apart from the extreme leading and trailing edges.  Only comp. II appears nearly unaltered in the ``abnormal'' mode, but also we note that its leading edge is now more completely polarized.

\begin{figure*} 
\includegraphics[width=146mm]{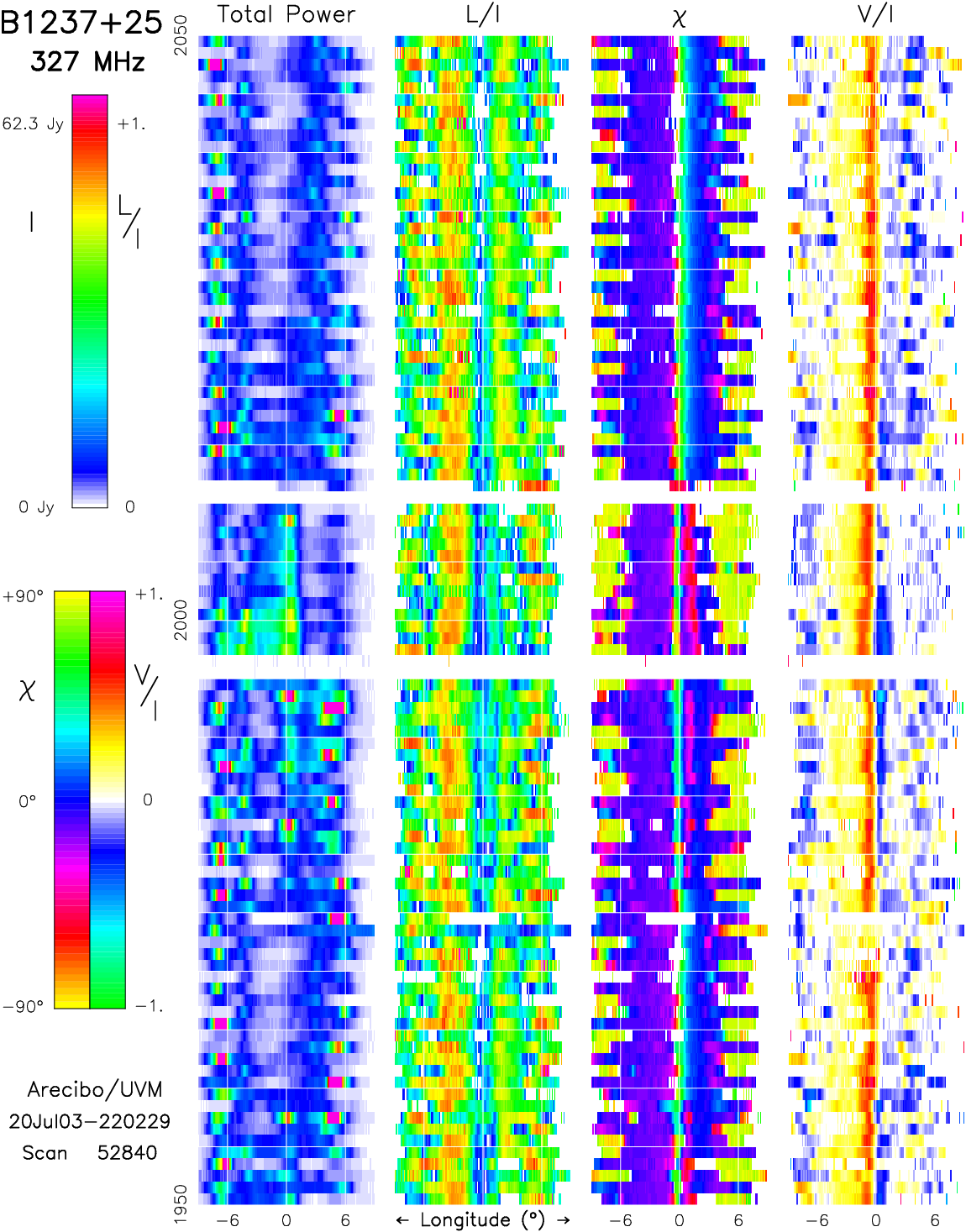}
\caption{Pulse-sequence polarization display showing three distinct behaviors: the ``normal'' mode (pulses 1960--1977 and 2015--2050), ``abnormal'' (1998 to 2010), and ``flare-normal'' mode (1978 to 1994) of the 2003 July 21 observation.  The total power $I$, fractional linear $L/I$, PA $\chi$, and fractional circular polarization $V/I$ are colour-coded in each of four columns according to their respective scales at the left of the diagram.  Core-component emission is virtually absent in ``quiet normal'' mode, the conal components exhibit regular modulation with a $P_3$ of 2.8 $P_1$, marked by a cyclic alternation of PPM ($-$10\degr, purple) and SPM (+80\degr, chartreuse) power on the outer profile edges.  The core component ``flares'' in the ``abnormal'' mode to almost continuous brightness, the conal modulation ceases, and SPM power predominates on the profile edges.  The ``flare-normal'' mode can be distinguished within the total-power column by the presence of core activity together with something of the ``normal'' conal modulation.  It also exhibits distinct linear PA and circular characteristics as compared with ``abnormal'' and ``quiet normal'' intervals.  Note the frequent presence of RH circular polarization in both ``flare-normal'' and ``abnormal'' sequences, but the two modes differ markedly in their PA signatures, especially around the core longitude.  The ``flare-normal''-mode PS here is longer than most and appears to exhibit a longer $P_3$ than in the proceeding ``quiet normal'' interval.  The PS between pulses 1950 and 1958 perhaps falls somewhere between the ``flare-normal'' and ``abnormal'' modes as it exhibits both the frequent core activity and conal modulation of the former, but shows a PA traverse more like the latter.  Finally, we see numerous individual pulses such as 1997 and 2012 where very partial emission within the pulse window is nonetheless polarized at an appropriate PA for that longitude.  Both the background noise level and interference level of these observations is exceptionally low with the latter effectively disappearing into the lowest intensity white portion of the $I$ color scale. }
\label{Fig5}
\end{figure*}

\section[]{Modal Intensity and Polarization Dynamics}
We have noticed that the three largely distinct modal behaviors described above can be distinguished by their polarization signatures at the individual pulse level---and indeed we find that they can be most accurately and reliably so identified.  Figure~\ref{Fig5} displays the full polarization state of a 100-pulse sequence of the 327-MHz observation of 2003 July 21.  The particular PS in Figure~\ref{Fig5} was chosen because, unusually, good examples all three modal behaviors fall within this short interval.  

\subsection[]{``Quiet-normal'' Mode}
This fundamental and most characteristic mode can be identified in total power by noting the intervals where the conal components exhibit their usual 2.8-$P_1$ modulation. This is most dramatically seen in the PA column, where the modal power on the profile outer edges alternates cyclically between some $-$10\degr\ (purple) and +80\degr\ (chartreuse). This outer-edge OPM modulation is also discussed for B1237+25 by \citet{b10}; see their fig. 4 for a 430-MHz example.  During these intervals the central core emission is weak or absent, though there is usually sufficient linear power in the profile center to define the PA.  Note that this PA behaves just as one would expect for a nearly central sightline traverse:  we see the $-$10\degr\ (purple-angled) linear polarization first rotate negatively toward $-$30\degr\ (magenta) and then nearly in the next sample become +60\degr\ (green) and then +30\degr\ (cyan)---corresponding to a $-$120\degr\ excursion---whereafter the PA rotates more gradually around to 0\degr\ (blue).  Components II and IV exhibit high levels of fractional linear power, with II frequently showing nearly complete polarization.  Finally, the fractional circular polarization peaks positively before and within the center of the profile, reaching some 60\% at about the position of the core component.  Note that the negative circular is much less consistent in this mode; at times there is weak RHC following the LHC peak, but often not as well.  We see little order to the circular polarization under the conal components in this mode.

\subsection[]{``Flare-normal'' Mode}
Primarily, the ``flare-normal'' mode differs from the ``quiet normal'' mode by the presence of bright core emission.  Note that this core activity is not continuous, but fluctuates from pulse to pulse in a manner that at times appears to mimic the conal modulation.  Though these ``flare-normal''-mode apparitions are characteristically of short (5--15 $P_1$) duration, they suggest (as in Fig.~\ref{Fig5}) a somewhat longer $P_3$, perhaps of about 4 $P_1$/c.  Note especially the contrasting PA modulation as compared to the ``quiet-normal'' mode.  We see little distinction between the ``normal'' modes in $L/I$, but there are clear differences in their PA and $V/I$ properties.  ``Flare-normal''-mode intervals exhibit a strange but consistent single-pulse PA behavior: at about comp. II's longitude we again find a $-$10\degr\ PA (purple-angled) which now first rotates {\em positively} through 0\degr\ (blue) to +30\degr\ (cyan), then reverses to rotate {\em negatively} through 0\degr\ (blue) to $-$30\degr\ (magenta), and then finally rotates {\em positively} again to 0\degr\ (blue).  A transition from ``quiet''- to ``flare-normal''-mode PA traverse properties is seen clearly at pulse 1978 and one in the opposite sense occurs at pulse 1960.  These PA sense reversals cannot be produced geometrically, so must have some other cause as we will discuss below.  Finally, note that the fractional circular polarization of the ``flare-normal'' mode tends to be somewhat smaller, but more fully antisymmetric than in the ``quiet normal'' mode.  Most of these circumstances are well demonstrated by modal averages as in Fig.~\ref{Fig2}, but we have found it necessary to carefully inspect the individual-pulse behavior to assess just how they come about.  

Overall, it is instructive to examine the ``quiet''- to ``flare-normal''-mode transition at about pulse 1978.  Note the pronounced narrowing of the emission window, particularly on the trailing side of the profile.  Comp. V becomes as bright as comp. I, but its linear polarization decreases---apparently as a result of mode-mixing because the individual-pulse fractional-linear distribution seems to change little across the boundary.  The RHC power under components I and V becomes more orderly, apparently contributing to the small residual seen in the modal profile.

\subsection[]{``Abnormal'' Mode}
``Abnormal''-mode intervals are easily distinguished from the other behaviors using only the total power. Not only does the core ``flare'', emitting almost continuously in successive pulses, but at meter wavelengths strong ``abnormal'' emission only occurs in the lefthand portion of the pulse window---and what remains of comps IV and V merge into a single weak ``hump''.  Moreover, all evidence of the usual modulation ceases and SPM power dominates over large leading and trailing portions of the modal profile.\footnote{Psaltis \& Seiradakis (1996) find evidence for weak conal modulation in the ``abnormal'' mode at 21 cms.; however, this might be explained by the inclusion of ``quiet-normal''--mode intervals, which could not be easily distinguished at this frequency.}  In $L/I$ we see little change in the ``abnormal'' mode, except at the longitude of comp. IV.  The PA and $V/I$ behaviour, however, show that the ``flare-normal'' and ``abnormal'' modes share common features, but also remain distinct.  Both the absence of modulation and SPM dominance are well known characteristics and immediately clear from the PA column of Fig.~\ref{Fig5}.  Note, however, that we again see the strange multiple PA reversals, but in the ``abnormal'' mode the PA tends to first rotate negatively (to $-$30\degr, magenta) as the core is approached, then jump positively (to +30\degr, cyan), rotate negatively at first steeply and then more gradually (to $-$60\degr, full red), and finally jump back to 0\degr\ (blue).  It is useful to contrast this ``abnormal'' PA traverse with that of the ``flare-normal'' mode:  the jump to a +30\degr\ angle (cyan) occurs at just the same longitude in the two modes, but the linear power at this point in ``abnormal'' PSs is very small---note that this point coincides with the $L$ minimum in Fig.~\ref{Fig4}.  The ``abnormal'' PA rotation under the core rotates further and much of it occurs at later longitudes.  

We note that the short ``abnormal''-mode PS of Fig.~\ref{Fig5} is not fully representative of the longer examples in our observations.  Comp. II is usually more consistently strong, so that we see a marked contrast between it and comp. IV in both the total power and fractional linear polarization.  Additionally, long ``abnormal'' mode PSs are typically more prominently circularly polarized with significant fractional LHC seen first under comp. II and rising to at least 60\% at the leading half-power point of the visible core.  At its trailing half-power point the RHC is typically only some 30\%, but it is important to emphasize that the LHC and RHC excursions represent about equal (antisymmetric) contributions of power.   

\begin{figure} 
\includegraphics[height=80mm,angle=-90.]{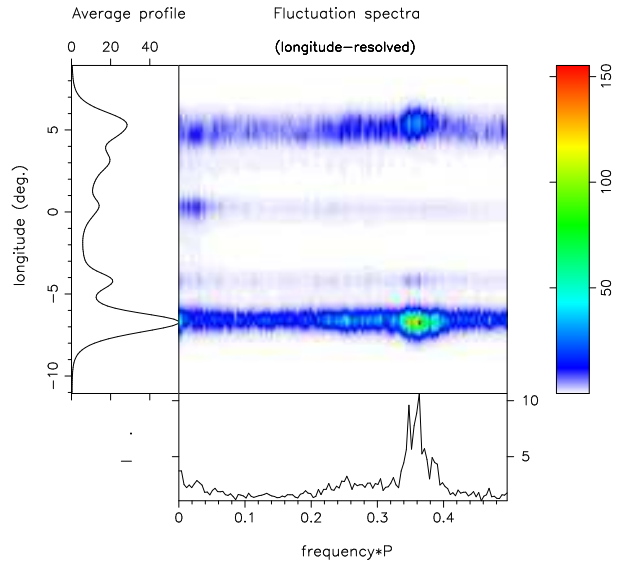}
\caption{Longitude-resolved, 256-point fluctuation spectra for the 5094-pulse, 327-MHz observation of 2003 July 14.  We see the usual bright pair of features near 0.35 c/$P_1$ representing modulation with a $P_3$ near 2.8 $P_1$, as well as weaker responses with a $P_3$ about 4 $P_1$.  Note that these features modulate components I, II and V, but apparently much less so IV.  Also note the low frequency features peaking near the longitude of the core component and modulating the broad center of the profile.  The larger peak here falls at some 0.027 c/$P_1$ or about 37 $P_1$, and the lower frequency peak may represent its subharmonic.  The spectra have been normalized by the total power at each longitude and corrected for interstellar scintillation by removing fluctuations having scales longer than some 1000 $P_1$.}
\label{Fig6}
\end{figure}

\subsection[]{``Normal''-mode Fluctuation Properties}
Figure~\ref{Fig6} gives longitude-resolved fluctuation spectra for our longest observation, the 5094-PS from 2003 July 14.  This PS exhibits little ``abnormal''-mode emission, thus it effectively represents the classical ``normal'' mode.  We see strong features near 0.35 c/$P_1$ as well as a broad weaker one around 0.25 c/$P_1$, corresponding to the usual $P_3$=2.8 $P_1$ as well perhaps as the nearly 4 $P_1$ modulation seen in the ``flare-normal'' mode.  

Furthermore, we also see the low frequency feature reported earlier[\citet{b15}, \citet{b16}, \citet{b2}, \citet{b9} ].  The brighter of the two features implies a $P_3$ of some 37 $P_1$ and the adjacent response corresponds to about twice this value.  One of us [\citet{b14}] had tended to associate this modulation with the core component, but we surely now see here that its effect is more general.  The feature modulates not only the core, but a broad region around it surely including comps. II \& IV.  We therefore conclude that the low frequency feature represents the amplitude modulation of the quasi-periodic ``flare-normal''-mode apparitions.  Visual inspection of the corresponding PS appears consistent with this interpretation.

\section[]{New Components?}
We have investigated whether \citet{b3} (hereafter G\&G) are correct in asserting that B1237+25 has two additional minor components, just leading and trailing the core component, which together constitute a smaller ``further in'' emission cone.  Their analysis is based on a ``window-threshold'' method [\citet{b4}], which itself remains somewhat controversial, because power is sought only in certain discrete longitude ``windows''.   It is worth noting that G\&G give some analytical details only about the leading component of the new pair---at --2.1\degr\ as opposed to +1.6\degr, relative to the core component---and our results pertain largely to it as well.  

G\&G applied their method to a 318-MHz, 1915-pulse, total-power observation having an apparently undetermined mix of modes.  A cursory examination of the total and modal profiles in Figs. 1--4 reiterates that the region leading the core by a little over 2\degr\ generally represents a minimum in profile power.  Only in the ``abnormal'' mode does this region emit at a level comparable to the other bright components.  We cannot learn, however, from the ``abnormal'' average in Fig.~\ref{Fig4} whether this power between components II and III represents a distinct component or has some other origin.  Using a version of G\&G's method, which accumulated only ``abnormal''-mode pulses having significant power levels in a window some 2\degr--2.5\degr\ before the core, we computed a partial modal profile.  Though it contained only 13 of some 91 ``abnormal''-mode pulses in our 2003 July 21 observation, it is almost identical to the full ``abnormal''-mode profile in Fig.~\ref{Fig4}.  The feature on the leading edge of the core in both profiles does not resolve into a separate component.  Even if we interpret it as a ``component'', it falls considerably later than the one in G\&G's fig. 2b, which peaks about halfway between components II and III.  Similar results were obtained for the ``flare-normal'' mode, with only 3 pulses qualifying of some 370, and nothing like a distinct new component appearing.  

\begin{figure} 
\includegraphics[width=80mm]{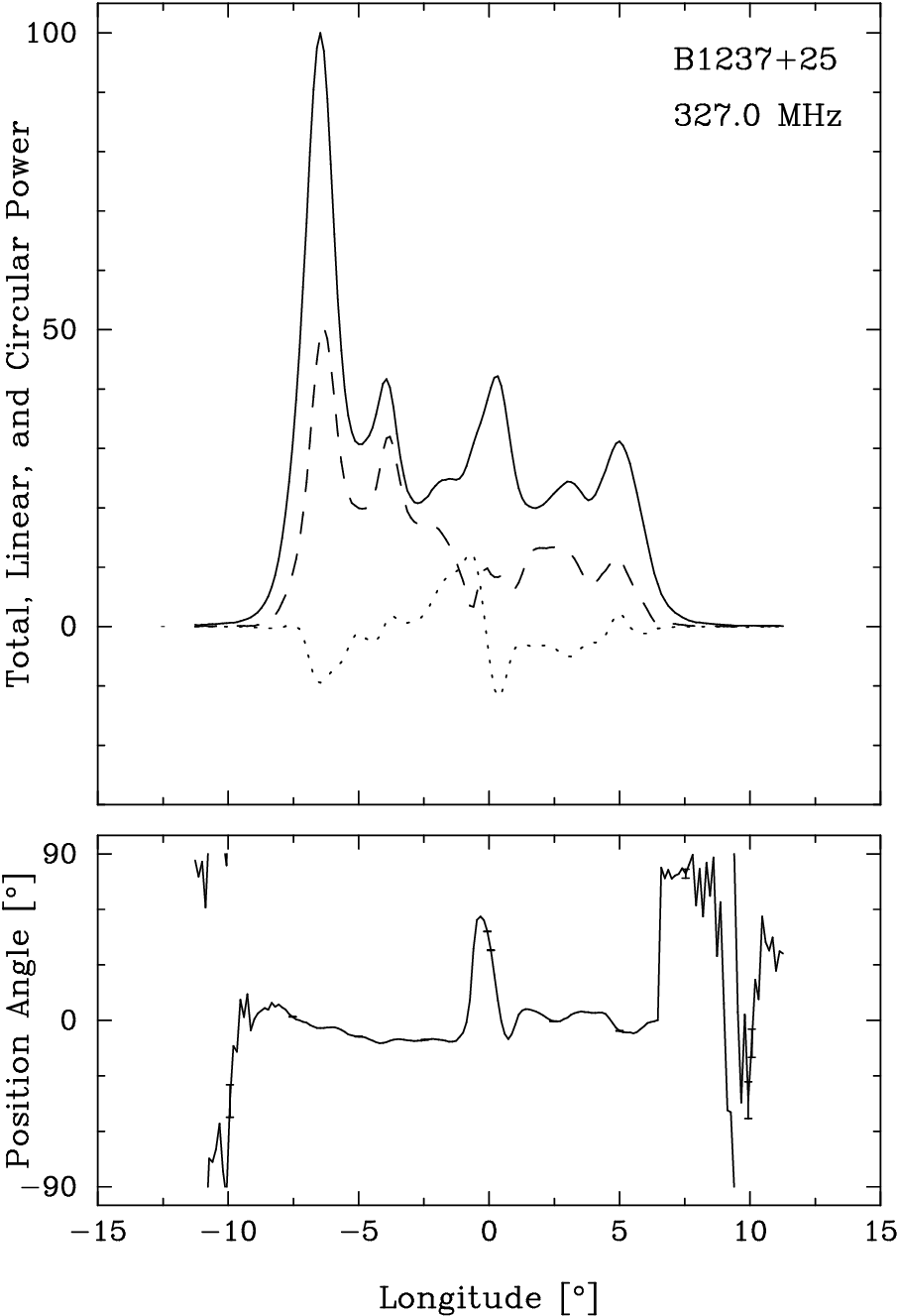}
\caption{Partial profile of the 31 qualifying pulses generated by applying a version of G\&G's ``window-threshold'' method to ``quiet-normal''-mode intervals (2654 pulses total) of the 2003 July 21 observation.  The selected single pulses have power greater than 108 times the off-pulse noise standard deviation in a ``window'' preceding the core-component peak by between 2.5 and 2.0\degr.   Note the new feature at just this longitude.  Note also, however, the surprisingly strong core component (given that the parent population were ``quiet normal'' pulses) together with the strong antisymmetric circularly polarized signature marking both the core and new feature.}
\label{Fig7}
\end{figure}

A different result was obtained when we applied G\&G's analysis to groups of ``quiet normal'' pulses, and this partial profile appears in Figure~\ref{Fig7}.   Here, we do see a discrete feature at about the same longitude shown in their fig. 2b.  Fig.~\ref{Fig7} averages 31 individual pulses which exceeded 108 times the off-pulse noise level $\sigma$---this out of a total of 2654 ``quiet normal'' pulses in the 2003 July 21 PS.   The feature is not so clearly resolved as in G\&G's fig. 2, but its position and upwardly slanting baseline suggest that we have succeeded in roughly verifying their published analysis.  

Please note that our Fig.~\ref{Fig7} differs markedly from G\&G's fig. 2.  Ours more resembles a hybrid of the star's ``flare-normal''-mode and total profiles, whereas theirs would seem to be dominated by ``flare-normal''  and ``abnormal''-mode power.  Perhaps the difference can be understood in terms of the greater sensitivity of the AO observations, where more weaker ``quiet normal''-mode pulses were found to exceed an appropriate threshold.  G\&G do not tell us how many pulses went into their partial profile, but both its shape and the narrowness of the ``new feature'' might be understood if it were very few.  

This said, the partial profile of Fig.~\ref{Fig7} is peculiar: we see that the ``new feature'' is accompanied by a relatively strong core---including its antisymmetric circular polarization---and that the ``new feature'' is as closely marked by the LHC as the core is by the RHC polarization.  Moreover, the relatively strong core and pronounced leading/trailing asymmetry are reminiscent of ``abnormal''-mode profiles, despite the fact that the qualifying pulses here were selected from a ``quiet-normal'' population having no clear  active-core intervals.  We will return to this finding below.  

Perhaps unfortunately, our similar effort to verify G\&G's purported trailing feature has met in failure. The core component (III) has a broad base in the two ``normal'' modes, as can readily be seen in the last several figures, and thus fills most of the ``gap'' with comp. IV.  Even in the ``abnormal'' mode, where this trailing region is relatively empty, it proved impossible to define the window so as to exclude power associated with the ``tail'' of the core.  Indeed, given the substantial width of the core component, we find it difficult to understand how these authors could have solved this problem. Their Table 3 gives the new component's position as +1.56\degr\ after the core peak, where comp. IV lies at 2.98\degr.  If the core is present, its trailing edge will surely overlie the position of the putative new component.  Only when the core is absent---or nearly so as in ``quiet normal''-mode PSs---would one seem to have a chance at finding this new component, and our searches for it in just this situation were negative as well.

\section[]{Nulling Analysis}
As reported previously, some 6\% of B1237+25's pulses are ``nulls'' [\citet{b34};\citet{b14}].  We are able to discriminate these nulls with reasonable accuracy using a threshold of about 10\% of the mean pulse intensity, depending on observation quality.  On this basis, null-length histograms show clearly that 1-pulse nulls are about 5 times more frequent than 2-pulse nulls, with the frequency declining steadily to a maximum of about 8--10 pulses.   The preponderance of 1-pulse nulls suggests that a population of unobserved and partial nulls also occur;  the former could be quite frequent (some 3\%), but the star's small duty cycle makes the latter rare (perhaps 1 in 1500 for this pulsar) and thus difficult to positively identify.  We have also looked for connections between nulls and modes, but nulls seem to occur within each of the modes and also at modal boundaries (\eg, see Fig.~\ref{Fig5}).  

One major hint, however, comes from constructing partial profiles comprised of the last pulses prior to a null or the first pulses afterward.  Overall, these partial-profile pairs strongly resemble each other and the total profile, and the differences (\eg strong comp. I after nulls) inconsistent among our various PSs.  One subtle change, however, is seen in all our 327-MHz observations:  the core component following nulls is elongated in the leading direction and often exhibits a bifurcated shape with the trailing part falling at the usual core position and the leading at just the longitude of the ``new component'' discussed in the preceding section.  There is also a tendency for this feature pair to be brighter just after nulls.  

We have thus investigated where the 31 qualifying pulses in Fig.~\ref{Fig7} fall and they appear mainly just before or after "flare-normal''-mode sequences, as identified from the PA as in Fig.~\ref{Fig5}.  This suggests that our mode-identification criteria may need to be extended to include such transitional pulses, given the "flare-normal'' mode total power and polarization characteristics of Fig.~\ref{Fig7}.  

One other interesting aspect of B1237+25's emission is a population of very weak pulses---surely qualifying as nulls according to the above noise threshold---which nonetheless exhibit polarized emission at angles appropriate for their longitude and position within the PS.  Two examples, seen in Fig.~\ref{Fig5}, are pulses 1997 and 2012 which happened to fall within the plotted interval, but other instances of the same effect are seen in any similar 100-pulse display.  This surely suggests that nulls represent a more complex phenomenon than the simple bi-state (on/off) received understanding, and we note with interest the recent paper by Janssen \& van Leeuwen (2004) which attributes a sort of ``sputtering'' to nulls in pulsar B0818--13.

\section[]{How Do the Three Modes Interact?}
We have not been able to see clearly whether transitions between the three modes exhibit any particular pattern or preferred sequence.  It is tempting to see the ``flare-normal'' mode as being intermediate between the extremes of ``quiet normal'' and ``abnormal'' behavior.  Both the ``flare-normal'' and ``abnormal'' modes have stronger core emission, somewhat narrower profiles, and a tendancy for components IV and V to merge.  The ``normal''-mode pair share the key property of regular subpulse modulation in the outer-cone regions (though perhaps the ``flare-normal'' mode has a somewhat longer $P_3$) as well as enough regularity in their succession to give the low frequency fluctuation-spectral feature.  One might also see the ``quiet normal'' mode as exhibiting the least SPM power, the ``flare-normal'' somewhat more, and the ``abnormal'' dominated by it. 

The low frequency modulation feature associated with``flare-normal''-mode apparitions argues that the two ``normal'' modes exhibit a cyclic relationship.  This conclusion is strengthened by the fact that neither mode persists for very long.  ``Flare-normal''-mode intervals rarely last more than a dozen pulses and ``quiet-normal'' sequences hardly 50.  The same cannot be said about the ``abnormal'' mode; it can switch on for a few pulses (as we see in Fig.~\ref{Fig5}), and our observations include a sequence 250 pulses long.  It would appear that the star does not often or easily change from the ``quiet normal'' to the ``abnormal'' mode.  As shown in Fig.~\ref{Fig5}, the more frequent transition is between ``flare-normal'' and ``quiet normal'', perhaps indicating that more drastic geometric or energetic changes are required for ``abnormal'' emission.  Once in the ``abnormal'' mode, however, it can stably continue much longer than the other modes.

\section[]{Core-component Width and Conal Emission Geometry}
The width of B1237+25's core component is of interest because of its close connection to the full angular size of the stellar polar cap, and therefore is an indicator of the star's emission geometry (Rankin 1990, 1993).  In many profiles the relative weakness of the core makes its width difficult to determine and, even when not so, measurements often give values smaller than the polar-cap size.  Added to this, the core is seen to substantially lag the center of the profile---that is, the midpoint between components I and V.  We will discuss both issues in turn.

Because B1237+25 exhibits a double cone and because we have presumed it to have a highly central sightline traverse (an issue we will verify below), its magnetic geometry can be computed reliably on the basis of its conal dimensions alone.  This computation was carried out in Rankin (1993) showing that the star's magnetic latitude $\alpha$ is some 53\degr.  This value agrees well with Lyne \& Manchester's (1988) somewhat different analysis giving 48\degr.  On this basis, we can then compute the 1-GHz core width expected from the polar-cap angular width as $W_{core}$=2.45\degr\ $P^{-1/2}/\sin\alpha$ [Rankin (1990); eq.(5)], and this in turn gives a width value of 2.61\degr.  This is rather more than is usually measured as can easily be verified from the foregoing profiles:  the total profile of Fig.~\ref{Fig1} shows the difficulty of determining the core width, but a generous estimate here might be 2.5\degr.  However, the isolated ``flare-normal'' and ``abnormal''-mode core components in Figs.~\ref{Fig2} \& \ref{Fig4} have widths of only some 1.55\degr--1.75\degr.  

The spectral behavior of the pulsar's core width is not fully known.  The relative weakness of its core and the narrower overall profile above 1 GHz make it increasingly difficult to measure this width.  At low frequencies some observations suggest that the core width is roughly independent of frequency, but most also suggest a width somewhat less than 2.6\degr\ [\eg \citet{b18}].  Indeed, the several published  observations in the 110--150-MHz range (Backer 1970; Lyne \etal\ 1971; Hankins \& Rankin 2005) all suggest a ``narrow'' core with a slow leading and steep trailing edge.  More illuminating are 112-MHz Pushchino profiles made in 2000, each 153 pulses long and with a resolution of about a milliperiod.  In 
some of these the core is nearly as bright as the other components, often shows a double form with a stronger following feature, and has an overall half-power width of hardly 2.8\degr\ (Suleymanova 2005).  This surely tends to reiterate that the B1237+25 core is "double" and that its width changes little, at least at meter wavelengths. 

At 327 MHz, the core shows a similar shape; either we see evidence for some leading-edge extension as in Figs.~\ref{Fig2} \& \ref{Fig4}, or for a leading-edge feature as in Fig.~\ref{Fig7}.  We are thus left with a consistent impression that the star's core is incomplete (or ``absorbed''; \eg \citet{b27}) on its leading edge.  This interpretation is further supported by the analyses of Qiao \etal\ (2000, 2003), who through Gaussian-component fitting to average profiles argue for a sixth weak component just prior to the visible core.  

\begin{figure} 
\includegraphics[width=80mm]{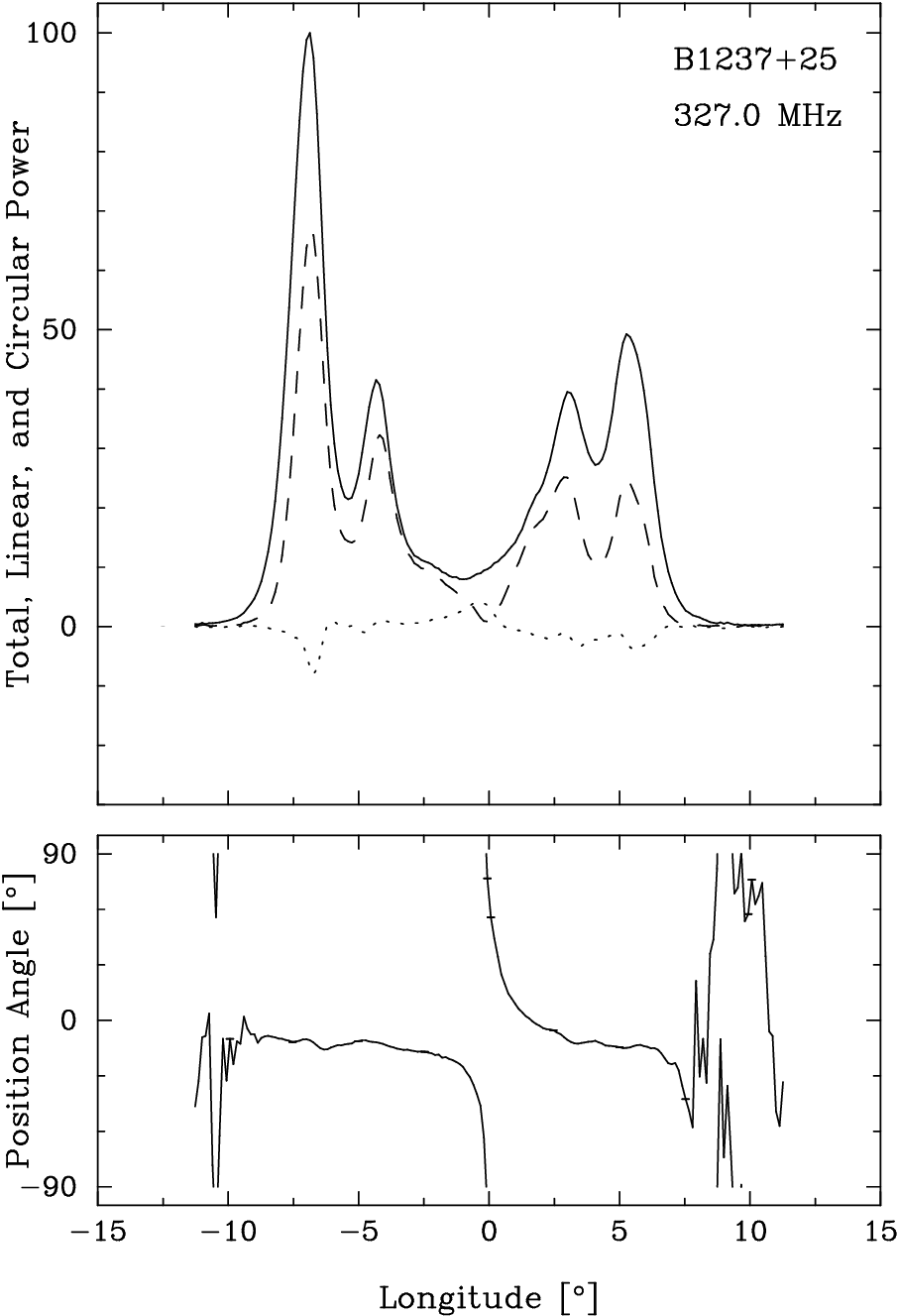}
\caption{``Quiet normal'' average profile comprised of 52 successive pulses wherein core activity was particular weak or absent.  This sequence was drawn from the 327-MHz observation on 2003 July 21 and includes pulses 704 through 755.  Note the almost total absence of a core component as well as the full negative, about 180\degr\ PA traverse.  Further note that the inflection point of the PA traverse falls within half a sample of the longitude origin; this convention was also used for each of the foregoing profiles.}
\label{Fig8}
\end{figure}

Additional strong support for this interpretation comes from the core's circularly polarized signature.  In the modal profiles above, we see that the visible core component aligns closely with the trailing negative (RHC) peak; whereas, the positive (LHC) peak falls in the ``extended'' region at about the half-power point on the core's leading edge.  This asymmetry in the total-power shape of the core component together with the balanced antisymmetric circular polarization is very unlike the typical core behavior in other pulsars.  Indeed, were we to take the ``core width'' not as the 3-db total power value, but rather as the interval between the respective leading and trailing half-power points on the circular signature, a value near 2.6\degr\ can be scaled from the profiles of Figs.~\ref{Fig2} \& \ref{Fig4}.  Or, to carry this one step further, we can measure the core's half width in ``flare-normal'' and ``abnormal''mode profiles between the circular zero-crossing point and the visible total-power component's trailing half-power point---and this gives values around 1.3\degr\ as expected.  

Finally, in Figure~\ref{Fig8} we give a modal profile of a ``quiet normal''-mode PS in which the core emission was particularly weak.  Note that here we see no core feature at all in either the total intensity or the circular polarization.  What we do see, however, is the full conal PA traverse!  Apart from the OPM dominance ``jumps'' at the outer edges of the profile, here finally is the expected PA traverse which is completely in keeping with the single-vector model [\citet{b23}; \citet{b24}]. We have taken the longitude origin in this and the earlier figures to fall within half a sample of the PA-traverse symmetry point.  Note that the PA traverse accumulates a full 180\degr\ in the longitude interval between components I and V and also that the PA rate near the center of the traverse is exceptionally steep as expected.  We measure the maximum PA-rate value to be some $-$185$\pm$5 \degr/\degr, which may well be the largest ever determined---though this may be an underestimate owing to averaging over pulses which ``jitter'' slightly in their longitude centers (see Fig.~\ref{Fig4}).  Using this value (rather than infinity) to compute the impact angle $\beta$ [see Rankin (1993), Table 2], we find that $\beta$ is 0.25\degr\ and $\beta/\rho$ 0.051, where $\rho$ is the outer cone radius to the outside 3-db point.

\section[]{Core-component Linear Polarization}
An old B1237+25 mystery is why its total profile fails to exhibit the expected 180\degr\ PA ``swing'' expected for a nearly central sightline traverse.  We have been able to show in Fig.~\ref{Fig8} that, indeed, the pulsar {\em does} exhibit such a traverse---but only in partial profiles restricted to the individual-pulse population having virtually no core emission.  This clearly established, we are now in a position to begin to understand what {\em core} characteristics are responsible for the distorted PA traverse we generally observe.  Or, said the other way around, this modal profile shows us how fragile is this full conal R\&C PA traverse, because it is carried by pulses which not only have little emission at the longitude of the core but are also linearly depolarized---probably because the confounding  core emission is never entirely absent.  

It then follows that when the core is present, both most of the power and most of the polarization belong to the core component.  This surely explains why the PA ``disruption'' is confined to a longitude region just under the core.  Note then that in those profiles above where core power is present, the PA exhibits a ``hook'' entailing usually {\em four} sense reversals.  (Only the ``quiet-normal''-mode profile in Fig.~\ref{Fig3} behaves differently---exhibiting two reversals---because core power is so weak here that the PA almost connects negatively between the two reversals.)  Then, we can observe that the PA centerpoint between the 2nd and 3rd reversals aligns accurately with the PA value far from the core (the average, for instance, under components 2 \& 4).  Such an unusual PA signature---which we have seen is encountered in both core-active modes---has a very natural explanation:  the core component must be dominated by linearly polarized power which is orthogonal to the conal power so clearly seen in Fig.~\ref{Fig8}.  And if this ``quiet normal''-mode conal power represents the primary polarization mode, it follows that the core is largely comprised of ``secondary'' polarization-mode emission.    

\begin{figure} 
\includegraphics[width=80mm]{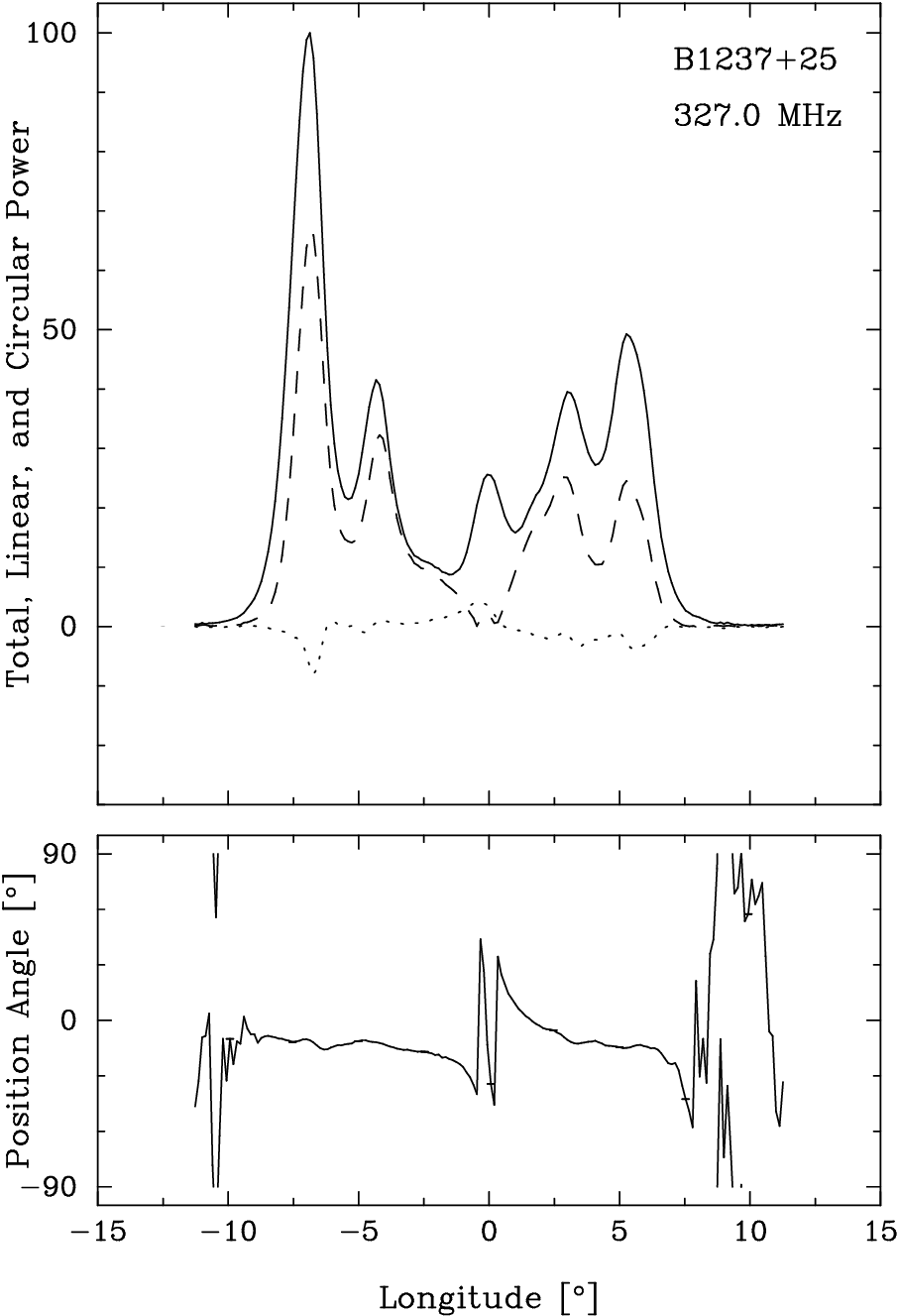}
\caption{Simulated, orthogonally polarized core component, together with the natural ``quiet-normal'' profile of Fig.~\ref{Fig8}. The Gaussian-shaped core component, centered at zero longitude was superposed on the natural profile. Note that the characteristic R\&C PA traverse is disrupted here in just the manner discussed above (see text)---that is, we see {\em four} PA sense reversals.  The simulated core is 20\% linearly polarized at a PA orthogonal to that of the natural profile in Fig.~\ref{Fig8}, has 20\% of the peak intensity and a $\sigma$ of 0.5\degr---though a broad variety of values produce the same qualitative result.}
\label{Fig9}
\end{figure}

This behavior can also clearly be seen in individual pulses.  In the context of discussing Fig.~\ref{Fig5} above, we pointed out the different PA behaviors exhibited by the three modes.   ``Quiet-normal''-mode pulses exhibit little evidence of PA reversals; whereas, both the ``flare-normal'' and ``abnormal'' modes show the same individual-pulse PA behavior as seen in their modal profiles.  Returning to these respective modal profiles in Figs.~\ref{Fig2} \& \ref{Fig4}, it is interesting to ponder the significance of their details.  In both cases, the centers of their PA ``hooks'' fall about 0.4\degr\ later than the zero-crossing points of their circular polarization signatures---points which nearly coincide with the center of the conal PA traverse in Fig.~\ref{Fig7}.  It is interesting then to see that the ``flare-normal''-mode profile shows no depolarized OPM-dominance boundary, whereas the ``abnormal'' profile exhibits one on the leading edge of the visible core component and perhaps one on its trailing edge as well.

Finally, Figure~\ref{Fig9} provides a simulation of the ``hook'' which we have found characteristic of the disrupted PA behavior under the core component.  We have simply added a Gaussian-shaped, linearly cross-polarized component to the natural core-absent profile of Fig.~\ref{Fig8}---and this results in the four-fold-reversed PA signature in the bottom panel.  This surely demonstrates that orthogonally polarized power must be a factor in both the depolarization and the nearly unique PA behavior seen in B1237+25.  It does not, however, fully delineate what the source of this cross-polarized power might be.  An obvious possibility is the SPM emission seen so prominently at other longitudes in this pulsar, but strictly SPM power is so far not known to be a property of core emission.  Were this the source, we might expect to see strongly depolarized mode-dominance boundaries, which are apparently seen in the ``abnormal''-mode profile of Fig.~\ref{Fig4}, but interestingly not so in the ``flare-normal''-mode profile of Fig.~\ref{Fig2}.  Emission-height variations such as suggested by Mitra \& Seiradakis (2003) may also provide an explanation for the observed polarization effects.  Detailed study of the depolarization will be required to distinguish these two alternatives.  

\begin{table*}
\centering
\caption{Aberration/Retardation Analysis Results for B1237+25}
\begin{tabular}{rcccccc}
\hline
      & $\phi^i_l$ & $\phi^i_t $ & $\nu^i$ & $\Gamma^i$ & $r^i_{em}$ & \\
  Cone   &  (deg) & (deg) & (deg)  & (deg) & (km) & $s^i_L$ \\
\hline
\hline
\\
wrt PA &&&&&&\\
outer & $-6.62\pm0.19$	&$5.43\pm0.16$&$-0.59\pm0.12$&$4.83\pm0.14$&$340\pm79$&$0.78\pm0.07$ \\
 inner & $-4.14\pm0.04$&$3.18\pm0.23$&$-0.48\pm0.12$&$2.94\pm0.13$&$278\pm76$&$0.53\pm0.05$ \\
\\
wrt CP &&&&&&\\
outer & $-6.52\pm0.19$&$5.53\pm0.16$&$-0.49\pm0.12$&$4.83\pm0.14$&$283\pm79$&$0.86\pm0.09$ \\
 inner & $-4.04\pm0.04$&$3.28\pm0.23$&$-0.38\pm0.12$&$2.94\pm0.13$&$220\pm75$&$0.59\pm0.07$ \\
\hline
\end{tabular}
\end{table*}







\section[]{Core-component Symmetry and Conal Emission Height}
Blaskiewicz, Cordes \& Wasserman (1991) and Malov \& Suleymanova (1998; hereafter M\&S) have attempted to account for core/cone asymmetries in pulsar emission profiles in terms of aberration and retardation effects and their analyses have led to useful methods for estimating the height of the emitting regions.  Both approaches require that the longitude positions of conal component pairs be measured relative to a fiducial longitude associated with that of the magnetic axis; and the two methods use, respectively, the linear PA-traverse and core-component centers, to estimate this position.  Extending their earlier work based on M\&S (Gangadhara \& Gupta 2001), G\&G applied their analysis to B1237+25.  As mentioned above, they reported a pair of new conal components comprising a third ``further in'' cone.  We, however, have only been able to verify their leading feature, which we find is clearly associated with the core component.  G\&G thus took the visible core component as indicative of the profile center, but we have seen above that only a trailing portion of the B1237+25 core is emitted.  For these reasons, their analysis using M\&S's technique is poorly founded for B1237+25 and thus their height estimates unreliable.  

In fact, our analysis above interestingly provides two possible fiducial points within the profile, the center (inflection point) of the linear PA traverse and the zero-crossing point of the antisymmetric circular polarization signature.  These two respective origins permit us to apply the methods of Blaskiewicz \etal\ (1991) and M\&S in the same star.  However, it is important to note that the two points do not occur simultaneously as one can be identified in ``quiet normal'' intervals and the other within ``flare-normal'' and ``abnormal'' PSs.  We give the measured (total or ``quiet normal''-mode) conal component positions relative to both the PA- and CP-defined origins in Table 1.  It is configured precisely as was G\&G's Table 3 and we also use their definitions for the conal leading and trailing component-pair positions $\phi^i_l$ and $\phi^i_t $, the total aberration-retardation longitude shift $\nu^i$, the magnetic azimuth angle $\Gamma^i$, emission height  $r^i_{em}$, and relative polar cap annulus $s^i_L$.  

We thus also base our computations on Gangadhara \& Gupta's (2001) eqs.(1-8, 10-15)] with one significant exception:  Dyks \etal\ (2004) have provided a subtle correction of their eq.(9) so we use the latter paper's eq.(7) instead---though in practice this correction results in only 10\% smaller emission heights for B1237+25.  Table 1 then shows that the outer and inner conal emission heights, relative to the PA traverse, are roughly the same, about 310$\pm$75 km, but that these cones are emitted on different field lines having their ``feet" at some 0.78$\pm$0.07 and 0.53$\pm$0.05 of the polar cap radius, respectively.  [These values are substantially smaller and more external than the ones computed by G\&G (2003)Ñthat is, 460 \& 600 km at 41 \& 59\%, respectively.]   The zero-crossing point of the circular signature was determined by making modal profiles of short  ``flare-normal'' or ``abnormal'' intervals, which were then compared with corresponding ``quiet normal'' intervals such as that shown in Fig.~\ref{Fig8}.  The circular zero-crossing point generally precedes the center of the PA traverse by some 0.10$\deg\pm$0.05$\deg$ (though we occasionally see the opposite behavior).  Computations based on this second definition of the profile center are also given in Table 1 where one can see there that the emission heights thereby obtained are some 60 km smaller and the annuli somewhat more exterior.  

We give this second calculation as an example, rather than an attempt to be definitive, as a thorough 
analysis and interpretation of the relation between the PA traverse and circular signature is beyond the scope of this paper.  On the scale of a few pulses with the large S/N and high resolution, the component shapes and positions fluctuate continuously, though some of the polarization variation appears to be systematic.  Note, for instance, the apparent ``drift'' to later longitudes of the positive circular during the last 50 or so pulses in Fig.~\ref{Fig5}.  Owing to these complexities, for example, we have not attempted here to interpret the narrowing of the profile in the core-active modes.  We plan to continue this work in a subsequent paper.

\section{Discussion and Conclusions}

This paper provides new---and we hope somewhat clarifying---analyses of the famous five-component ({\bf M}) pulsar B1237+25, based largely on the straightforward method of partial polarization profiles.  Pulsars with such clear double-conal/core structure remain unexplained by pulsar theories.  Not only has core emission thus far received little detailed analytical attention, but the properties of double cones also remain understudied.  Thus this pulsar still provides a primary context for such investigation; not only is it perhaps the brightest of the {\bf M} stars, but it also exhibits a nearly precise central sightline geometry, regular subpulse modulation, modes, OPM effects and nulls.  

The study began with a reexamination of the star's modes, subpulse modulation and nulls.  In addition to studying the star's ``abnormal'' mode, we find that its long known ``normal'' mode in fact consists of two distinct behaviors, which we have designated the  ``quiet normal'', where little or no core activity can be distinguished in individual pulses, and the ``flare-normal'' mode, where the core is nearly as bright as in the ``abnormal'' mode, but displays distinct properties.  In many ways this ``flare-normal'' mode can be regarded as an intermediate state between the ``quiet normal'' and ``abnormal'' behaviors.  In this ``flare-normal'' mode, the core is bright, but not so fully dominated by the SPM.  The profile narrows and components IV and V partially merge, but not as much as in the ``abnormal'' mode.  Subpulse modulation in the ``flare-normal'' mode is not entirely quenched (as in the ``abnormal''), but exhibits less regularity and perhaps a $P_3$ value around 4 $P_1$/c.  One can also view the ``flare-normal'' mode as exhibiting much more SPM than the ``quiet normal'' mode, but far less than the ``abnormal'' mode wherein it is dominant over most of the profile.  

``Flare-normal'' and ``quiet normal'' intervals alternate with each other quasi-periodically.  It is this regular appearance of bright ``flare-normal''-mode core power in the center of the profile that produces the long known ``core''-associated fluctuation feature (\eg Rankin 1986).  Typically, ``quiet normal'' sequences persist for some 40--70 pulses and ``flare-normal'' intervals for 5--15 pulses, making a complete cycle of some 60--80 pulses in duration.  We note that what appear to be short ``abnormal''-mode intervals are often interspersed within this overall ``quiet normal''/``flare-normal''-mode cycle;  generally, however, these persist for only a few pulses.  At unpredictable times, though, the ``abnormal'' mode appears to ``catch'', and then it can persist without interruption for a few or many hundreds of pulses.  ``Abnormal''-mode sequences thus appear to reflect a distinct relatively stable``state'' of the star's magnetospheric emission; whereas, the ``normal''-mode alternation of ``quiet'' and ``flare'' intervals represents a further cyclic ``state''.  

While it has almost uniformly been assumed that B1237+25 had a highly central sightline geometry, its average PA traverse does not at all simply bear this out.  Its total profile exhibits neither the constant PA nor the steep, 180$\deg$, central traverse expected for such a sightline geometry.  Some ``abnormal''-mode profiles (\eg Bartel \etal\ 1982) exhibited more of the expected PA traverse, but it remained to be explained why the total profile did not.  We have largely been able to resolve this issue.  The most core-quiet intervals of ``quiet normal''-mode PSs are found to exhibit an almost textbook-quality PA behavior, as shown above in Fig.~\ref{Fig8}.  The center of this PA traverse has a sweep rate of at least 180$\deg/\deg$ and the PA in the ``wings'' of the profile becomes almost constant at about the same PA value (apart from OPM dominance effects on the extreme profile edges.)  This, in turn, permits us to determine the star's $\beta$ value from its PA sweep rate for the first time, confirming that $\alpha$ is 53$\deg\pm$2$\deg$ (Rankin 1993) and $\beta$ 0.25$\deg\pm$0.05$\deg$.   In terms of the geometrical models of the above paper, this implies that $\beta/\rho$ is hardly 5\%.  

Another mystery has revolved around the behavior of B1237+25's core component.  First, its expected antisymmetric circularly polarized signature is weak at best in the total profile (\eg Fig.~\ref{Fig1}), and 
second, the observed width of its core component at meter wavelengths is often or usually substantially less than the expected 2.45$\deg P^{-1/2}/\sin\alpha$ (Rankin 1990).  Our 327-MHz ``abnormal''- (Fig.~\ref{Fig4}) and ``flare-normal''-mode (Fig.~\ref{Fig2}) profiles show beautiful anti-symmetric circular signatures under the core component.  However, it is noteworthy that this circular symmetry is retained despite the partial character of the total-power core component itself.  Thus we see that the leading portion of the core is at least 50\% circularly polarized, whereas the trailing portion (after the core peak) exhibits only 25--30\% fractional circular polarization.  We note also that the circular signatures in the ``abnormal''- and ``flare-normal''-mode profiles are nearly identical---that is, the two modal core components have about the same fractional circular polarization and both their total intensities and circular zero-crossings nearly overlie each other.  The strong difference in their respective core linear polarization is then noteworthy:  The linear peak falls late in the ``abnormal'' mode and is bounded on both sides by near minima; whereas, in the ``flare-normal'' mode the linear peak lies near the center of the component and the linear polarization appears continuous almost across the entire profile.  This difference can also be observed in the two PA traverses:  The ``abnormal''-mode ``hook'' under the core component falls later within the component and rotates further negatively, as we have also seen in individual pulses in Fig.~\ref{Fig5} above.  

We reemphasize that the B1237+25 core component is dominated by cross-polarized power.  The extent and manner in which this occurs varies from mode to mode.  Even in ``quiet normal'' PSs (\eg Figs.~\ref{Fig3} \& \ref{Fig5}) the pronounced depolarization under the core appears to corroborate the contribution of orthogonally polarized power.  Also, the relative contribution of SPM power becomes stronger in the ``flare-normal'' and ``abnormal'' modes, as can be seen very clearly in the colour PS polarisation display above.  Note in particular the manner in which the SPM power can be seen in the PA column on the outer edges of the emission window.  We have been able to model the PA ``hook'' by adding cross-polarized power in the form of the core to the ``quiet normal'' profile.  We cannot yet be sure whether the cross-polarized power in the core is SPM power or, for instance, it results from differences in emission height as suggested by Mitra \& Seiradakis (2003).

The width of the B1237+25 core component at meter wavelengths remains an issue.  We find that the visible core in modal profiles is hardly 2$\deg$, substantially less than the expected at least 2.6$\deg$.  However, this visible core is clearly late, as it aligns with the trailing, RHC peak of the antisymmetric circular signature.  This trailing portion of the core seems to be complete as its width, measured between the circular zero-crossing and its trailing 3-db points, is just half of 2.6\degr\ as expected.  We see little which aligns with the leading LHC peak in the modal profiles.  However, this is just the point where G\&G found the earlier of their two ``new components'' using their window-threshold method---and this is the feature which we have been able to verify as well using their method.  G\&G aside, we do occasionally see individual pulses which have peaks at the longitude of the LHC peak, about --1$\deg$.  

We therefore conclude on the basis of the various evidence that the core component in B1237+25 is asymmetric and incomplete.  However, we find evidence for a complete core---with about the expected 2.5--3$\deg$ width---in both the scale of the two respective peaks of the circular signature and also by the observed emission in some single pulses which align with the LHC peak.  Some short PSs at lower frequencies seem to exhibit just this sort of core more clearly (Suleymanova 2005).  It is then hardly surprising that when Qiao \etal\ (2003) fitted Gaussian functions to this star's profile, they found it necessary to add a sixth component at just this ``leading core'' longitude.  They also concluded that the core is hollow and we find no clear evidence from our work to contravene this conclusion---though neither can we fully verify it.  B1237+25 is nearly unique in providing a traverse so close to the magnetic axis.  

Also, we used the unusually clear definition of both the PA-traverse and circular symmetry points that our analysis provides to estimate the star's emission heights using the recent reformulation of Dyks \etal\ (2004).  We find respective outer and inner emission-cone heights of 340$\pm$79 km and 278$\pm$76 km relative to the PA-traverse inflection point, such that they lie on polar cap annuli of 78 and 53\%.  We also determine these heights with respect to the circular polarization zero-crossing point giving us values some 60 km smaller.  An obvious interpretation of this difference might be that the core emission is emitted much closer to the stellar surface, but at a height of some 60 km.     

Finally, although our study is focussed primarily on core characteristics. it is interesting to compare our results which those of Psaltis \& Seiradakis (1996).  We are not surprised that they found evidence for three emission rings in total power, and we believe that these should be associated with the inner cone as well as the PPM and SPM subrings which comprise the outer cone (Rankin \& Ramachandran 2003).
Our results further tend to support their conclusion that the conal modulation feature is comprised of several preferred values (see Fig.~\ref{fig6}), and it is surely true that the low frequency feature is associated more with the inner conal components than the outer ones.  Our present analysis has not been aimed at discerning the subbeam ``carousel'' structure, but it is difficult for us to understand how the low frequency ``modal'' modulation---in which the core participates---could be commensurate with the well known 2.8-$P_1$ modulation seen prominently in the outer components.

Surely, our new analyses of B1237+25 above have demonstrated that this marvelous star still has very much to teach us.  For almost no other pulsar do we have the opportunity to study the characteristics of the emission so close to the magnetic axis---and the pulsar remains the paragon of the five-component ({\bf M}) species.  We plan to continue some of the lines of investigation reported here and, in particular, expect to report further on the star's core cross-polarized emission in a subsequent paper.

\section*{Acknowledgments}
We thank Avinash Deshpande for help with the observations, Stephen Redman and Svetlana Suleymanova for assistance with aspects of our analysis, and Jaroslaw Dyks, Svetlana Suleymanova, and Geoff Wright for discussions.  Portions of this work were carried out with support from US National Science Foundation Grants AST 99-87654 and 00-98685.  Arecibo Observatory is operated by Cornell University under contract to the NSF.  This work made use of the NASA ADS system.

\bsp

\label{lastpage}


\begin{thebibliography}{99}
\bibitem[\protect\citeauthoryear{Backer}{1970a}]{b22a} Backer, D.C. 1970a, Nature, 228, 42 
\bibitem[\protect\citeauthoryear{Backer}{1970b}]{b22b} Backer, D.C. 1970b, Nature, 228, 752 
\bibitem[\protect\citeauthoryear{Backer}{1970c}]{b22c} Backer, D.C. 1970c, Nature, 228, 1297
\bibitem[\protect\citeauthoryear{Backer}{1973}]{b2} Backer D.C. 1973, Ap.J.,182, 245
\bibitem[\protect\citeauthoryear{Backer}{1976}]{b13} Backer D.C. 1976, Ap.J.,209, 895
\bibitem[\protect\citeauthoryear{Backer \& Rankin}{1980}]{b12} Backer D.C., \& Rankin, J.M. 1980, Ap.J.Suppl., 42, 143
\bibitem[\protect\citeauthoryear{Bartel et al.}{1982}]{b9} Bartel N., Morris D., Sieber W., Hankins T.H. 1982, Ap.J., 258, 776
\bibitem[\protect\citeauthoryear{Blaskiewicz, Cordes \& Wassermann}{1991}]{b26} Blaskiewcz, M., Cordes, J.M., \& Wassermann, I. 1991 Ap.J., 370 643
\bibitem[\protect\citeauthoryear{Dyks \etal\ }{2004}]{b28} Dyks, J., Rudak, B., \& Harding, A.K. 2004, Ap.J., 607, 939
\bibitem[\protect\citeauthoryear{Gangadhara \& Gupta}{2001}]{b4} Gangadhara R.T., Gupta Y. 2001,
Ap.J., 555, 31
\bibitem[\protect\citeauthoryear{Gupta \& Gangadhara}{2003}]{b3} Gupta Y., Gangadhara R.T. 2003,
Ap.J., 584, 418
\bibitem[\protect\citeauthoryear{Hankins \& Wright}{1980}]{b1} Hankins T.H., \& Wright G.A.E. 1980,
Nat, 288, 681
\bibitem[\protect\citeauthoryear{Hankins \& Rankin}{2005}]{b18} Hankins T.H., \& Rankin, J.M. 2005, A.J., preprint
\bibitem[\protect\citeauthoryear{Janssen \& van Leeuwen}{2004}]{b8} Janssen, G.H., \& van Leeuwen, A.G.L. 2004, Astr. \& Astrop., 425, 255
\bibitem[\protect\citeauthoryear{Komesaroff}{1970}]{b24} Komesaroff, M.M. 1970, Nature, 225, 612
\bibitem[\protect\citeauthoryear{Lyne, Smith \& Graham}{1971}]{b32} Smith, F.G., \& Graham, D.A. 1971, M.N.R.A.S., 153, 337
\bibitem[\protect\citeauthoryear{Lyne \& Manchester}{1988}]{b20} Lyne, A.G., \& Manchester, R.N. 1988, M.N.R.A.S., 234, 477
\bibitem[\protect\citeauthoryear{Malov \& Suleymanova}{1998}]{b25} Malov, I. F., \& Suleymanova, S.A. 1998, Astron. Rep. 42, 388
\bibitem[\protect\citeauthoryear{Mitra \& Seiradakis}{2003}]{b30} Mitra, D.  \& Seiradakis, J. M.  2003, Proceeding of the 6th Hellenic Astronomical Conference, Penteli, Athens, Greece, 15-17 September 2003, P. G. Laskarides, ed., p. 205
\bibitem[\protect\citeauthoryear{Oster \& Sieber}{1977}]{b6} Oster L., Sieber W. 1977, A\&A, 58, 303
\bibitem[\protect\citeauthoryear{Pr\'{o}szy\'{n}ski \& Wolszczan}{1986}]{b7} Pr\'{o}szy\'{n}ski M., Wolszczan A. 1986, Ap.J., 307, 540
\bibitem[\protect\citeauthoryear{Psaltis \&Seiradakis}{1996}]{b33} Psaltis, D., \& Seiradakis, J.M., 1196, Proceeding of the 2nd Hellenic Astronomical Conference, Thessaloniki, Greece, 29 June--1 July 1995, M.E. Contadakis, J.D. Hadjidemetriou, L.N. Mavridis, \& J.H. Seiradakis, eds., p. 316.
\bibitem[\protect\citeauthoryear{Qiao \etal\ }{2000}]{b21} Qiao, G.J., Liu, J.F., Wang, Y., Wu, X.J., \& Han, J.L. 2000, Pulsar Astronomy --- 2000 and Beyond (IAU Colloquium \#177, Bonn, Germany; M. Kramer, N. Wex \& R. Wielebinski, eds.) ASP Conference Series, 2002, 197
\bibitem[\protect\citeauthoryear{Qiao \etal\ }{2003}]{b17} Qiao, G.J., Li, K.J., Wang, H.G., Xu, X.Xu., \& Liu, J.F. 2003, Acta Astr. Sinica, 44, 230
\bibitem[\protect\citeauthoryear{Radhakrishnan \&Cooke}{1969}]{b23} Radhakrishnan, V., \& Cooke, D.J. 1969, Ap. Lett, 3, 225
\bibitem[\protect\citeauthoryear{Rankin }{1983}]{b27} Rankin J.M. 1983, Ap.J, 274 333
\bibitem[\protect\citeauthoryear{Rankin}{1986}]{b14} Rankin J.M. 1986, Ap.J., 301, 901
\bibitem[\protect\citeauthoryear{Rankin }{1990}]{b19} Rankin, J.M. 1990, Ap.J., 352, 247
\bibitem[\protect\citeauthoryear{Rankin}{1993}]{b11} Rankin J.M. 1993, Ap.J., 405, 285 and Ap.J.Suppl., 85, 145
\bibitem[\protect\citeauthoryear{Rankin \& Ramachandran}{2003}]{b10} Rankin J.M., Ramachandran R. 2003, Ap.J., 590, 411
\bibitem[\protect\citeauthoryear{Ritchings}{1976}]{b34} Ritchings, R.T. 1976, M. N.R.A.S., 176, 249
\bibitem[\protect\citeauthoryear{Stinebring \etal\ }{1984}]{b5} Stinebring D.R., Cordes J.M., Rankin J.M., Weisberg J.M., Boriakoff V. 1984, Ap.J.Suppl., 55, 247
\bibitem[\protect\citeauthoryear{Suleymanova}{2005}]{b31} Suleymanova, S.A. 2005, private communication.  
\bibitem[\protect\citeauthoryear{Taylor \& Huguenin}{1971}]{b15} Taylor, J.H., \& Huguenin, G.R. 1971, Ap.J., 167, 273
\bibitem[\protect\citeauthoryear{Taylor, \etal\ }{1975}]{b16} Taylor, J.H., Manchester, R.N., \& Huguenin, G.R. 1975, Ap.J., 195, 513
\end{thebibliography}
\end{document}